\documentclass[11pt]{amsart}
\usepackage{geometry}                
\input epsf
\usepackage{bm}
\usepackage{amsmath}
\usepackage{graphicx}
\usepackage{gensymb}

\usepackage{bm}
\usepackage{amsmath}
\usepackage{graphicx}
\usepackage{chemarr}
\usepackage{mathtools}
\usepackage{eufrak}

\usepackage{hyperref}

\hypersetup{
    colorlinks=true, 
    linktoc=all,        
    linkcolor=black,  
    citecolor=blue
}

\title{Topology by Design in Magnetic nano--Materials: Artificial Spin Ice}
\author{Cristiano Nisoli, Theoretical Division, Los Alamos National Laboratory, Los Alamos, NM, 87545, USA}

\date{\today}                                           

\begin{document}
\maketitle

\tableofcontents

{\bf Abstract}

Artificial Spin Ices are two dimensional arrays of magnetic, interacting nano-structures whose geometry can be chosen at will, and whose elementary degrees of freedom can be characterized directly. They were introduced at first to study frustration in a controllable setting, to mimic the behavior of spin ice rare earth pyrochlores, but at more useful temperature and field ranges and with direct characterization, and to provide practical implementation to celebrated, exactly solvable models of statistical mechanics previously devised to gain an understanding of degenerate ensembles with residual entropy. With the evolution of nano--fabrication and of experimental protocols it is now possible to characterize the material in real-time, real-space, and to realize virtually any geometry, for direct control over the collective dynamics. This has recently opened a path toward the deliberate design of novel, exotic states, not found in natural materials, and often characterized by topological properties. Without any pretense of exhaustiveness,  we will provide an introduction to the material, the early works, and then, by reporting on more recent results, we will proceed to  describe the new direction, which includes the design of desired topological states and their implications to kinetics.

\section{Introduction}

From quasi-particles, fractionalization, pattern formation, to the nano--machinery of life in DNA replication and transcription, or to the coherent behavior of a flock, an ant colony, or a human group, emergent phenomena are generated by the collective dynamics of surprisingly simple interacting building blocks. Indeed, much of the more resent research in condensed matter pertains to the modeling of unusual emergent behaviors, typically from correlated building blocks in natural materials, either at the quantum or classical level~\cite{anderson1972more}.  A few years ago~\cite{Wang2006} we proposed a different approach: design, rather than simply deduce, collective behaviors, through the interaction of simple, artificial building blocks whose interaction could lead to exotic states not seen in natural materials. 

Arrays of  elongated, mutually interacting, single-domain, magnetic nano-islands arranged along a variety of different geometries, (Fig. 1) were ideal candidates. The magnetic state of each island could be described by a classical Ising spin, and advances in lithography  allowed their nano--fabrication  in virtually any geometry. The advantage of such approach is twofold: (1) the low energy dynamics, which underlies possible exotic states, is dictated by geometry, which here is open to design; (2) characterization methods---Magnetic Force Microscopy (MFM), PhotoElectron Emission Microscopy (PEEM), Transmission Electron Microscopy (TEM), Surface Magneto-Optic Kerr Effect (MOKE), Lorentz Microscopy---allow direct visualization of the magnetic degrees of freedom for unprecedented validation. 
Nano-scale is an good choice choice: the size of the building blocks, which are shape-anisotropic, elongated nano--islands (typically, NiFe alloys $200\times80\times5-30$ nm$^3$ patterned by nano-lithography on a non-magnetic Si substrate), has to be inferior to the typical magnetic domain, to provide single domains with magnetization directed along the principal axis, that can be interpreted as  switchable spins.  

These Artificial Spin Ices (ASI) were employed  at first to study frustration in a controllable setting, to mimic the behavior of spin ice rare earth pyrochlores, but at more useful temperature and field ranges and with direct characterization, and to provide practical implementation to celebrated, exactly solvable models of statistical mechanics previously devised to gain an understanding of degenerate ensembles. Soon, a  growing number of groups has extended the use of ASI~\cite{Nisoli2013colloquium}, to investigate topological defects and dynamics of magnetic charges~\cite{Mengotti2010,Ladak2010,ladak2011monopole,zeissler2013non,phatak2011nanoscale,ladak2011direct,pollard2012propagation}, information encoding~\cite{Lammert2010,wang2016rewritable}, in and out of equilibrium thermodynamics~\cite{Nisoli2007, Nisoli2010,Ke2008,cugliandolo2017artificial,levis2013thermal,Morgan2010,budrikis2012disorder,budrikis2011diversity,Lammert2012,Nisoli2012}, avalanches~\cite{hugli2012emergent,mellado2010dynamics}, direct realizations of the Ising system~\cite{zhang2012perpendicular,arnalds2016new,nisoli2016nano}, magnetoresistance and the Hall effect~\cite{Branford2012,PhysRevB.95.060405}, critical slowing down~\cite{anghinolfi2015thermodynamic}, dislocations~\cite{drisko2017topological}, spin wave excitations~\cite{gliga2013spectral}, and memory effects~\cite{gilbert2015direct,libal2012hysteresis}. Meanwhile similar strategies~\cite{Libal2006, Libal2009,libal2016dynamic,Olson2012,Ray2013,nisoli2014dumping} have found realization in trapped colloids~\cite{ortiz2016engineering,tierno2016geometric,loehr2016defect}, vortices in nano-patterned superconductors~\cite{Latimer2013,Trastoy2013freezing} and even at the macroscale~\cite{Mellado2012}.  With the evolution of nano--fabrication and of experimental protocols it is now possible to characterize the material in real-time, real-space~\cite{Kapaklis2012,arnalds2012thermalized,Farhan2013,Porro2013,kapaklis2014thermal}, and to realize virtually any geometry, for direct control over the collective dynamics. This has recently opened a path toward the deliberate design of novel, exotic states~\cite{Morrison2013,Chern2013, gilbert2014emergent, gilbert2016emergent,stamps2014artificial} not found in natural materials~\cite{gilbert2016frustration,nisoli2017deliberate}.


\begin{figure}[t!]
\includegraphics[width=.9\columnwidth]{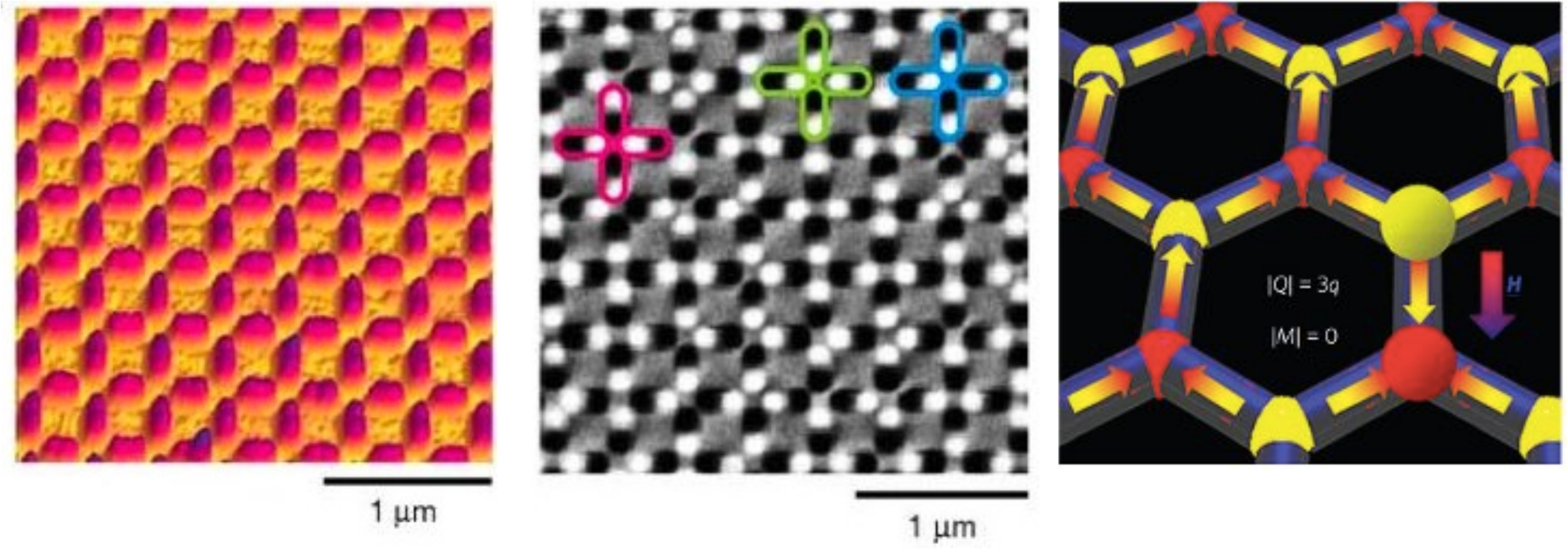}

\begin{center}
\caption{Artificial spin ice in its most common geometries. Left: Atomic force microscopy image of square ice showing its structure (figures from~\cite{Wang2006}). Center: a magnetic force microscopy image of square ice, showing the orientation of the islands' magnetic moments (north poles in black, south poles in white); Type-I (pink) Type-II (blue) and Type-III (green) vertices are highlighted (see the text and Fig.~7 for a definition).  Right: schematics of honeycomb spin ice.}
\label{default}
\end{center}
\end{figure}

Frustration is a fundamental ingredient in design: it controls the interplay of length and energy scales, dictating the emergent dynamical properties that lie at the boundaries between order and disorder, and leading to a lively, quasi-disordered ensemble called {\it ice manifold}, to be exploited in the design of exotic behaviors.

Correlated spin systems have of course a long history in Physics. In classical statistical mechanics, the Ising model~\cite{ising1925beitrag} paved the way to our understanding of long-range order from symmetry breaking as a second order phase transition, universality classes and scaling~\cite{kadanoff1993scaling}, and finally the renormalization group~\cite{wilson1971renormalization} with implications reaching well beyond condensed mater systems~\cite{wilson1974renormalization}. However, frustrated spin systems often do not order,  generally resulting in quasi-disordered manifolds governed by some geometric or topological rule. Often their collective dynamics lends itself to emergent descriptions that are only partially reminiscent of the constitutive spin structure. The situation is somehow similar to everyday life, where frustration results from a set of constraints that cannot be all satisfied at the same time, leading to a manifold of compromises among which the choice is most often equivalent and can be influenced by a small bias. Thus, obstructed  optimization provides high susceptibility that can generate the complex social dynamics we witness everyday. These analogies between social settings and frustrated materials  are not merely philosophical: ideas borrowed from the frustrated spin ice physics have been exploited in social networks to describe wealth allocation~\cite{mahault2017emergent}.

Much as in life, frustration is understood in Physics as a set of constraints that cannot be all satisfied. Typically the constraint is the optimization of an energy, usually the pairwise interaction between elementary degrees of freedom. This too  leads to a degenerate manifold which preserves non-zero entropy density at low temperatures, in apparent violation of the third law of thermodynamics.


We will see how frustration is exploited in the design of artificial spin ices.  Initially, the aim was  pure exploratory science, with the goal to understand frustration and disorder in a controllable environment that could be characterized directly. These materials could mimic the frustrated ice rule that defines the exotic manifold of rare earth pyrochlores (see below), yet at room temperature rather than at  the Kelvin scale. They could also provide the first realization of the celebrated exactly solvable models of statistical mechanics, such as the antiferromagnetic Ising system on a triangular lattice described above, or the various vertex models introduced and/or solved by Lieb, Wu and Baxter between the late 60s and early 80s~\cite{lieb1967residual,lieb1967exact,Wu1969,Baxter1981}. As both experimental protocols and theoretical understanding evolved, however, it became clear how the material could open new paths in a material-by-design effort: instead of finding, more or less serendipitously, natural materials of interesting or novel behavior, one could think about a bottom-up approach, where a suitable design could produce desired exotic properties. 

In this chapter we will start with a brief description of fundamental concepts and then earlier realizations, fleshing out the basics pertaining to their experimental protocols, nano-fabrication, and characterization (for details we refer to references or to the following reviews~\cite{Nisoli2013colloquium,marrows2016experimental,heyderman2013artificial}. Geometric frustration should really be called topological, as it is essentially a topological property, yet it is based on interaction, which is instead geometric, and generally not topologically invariant: we will discuss the two issues in parallel, and see how new approaches and  new designs, based on a different level of frustration, so called vertex-frustration,  can indeed allow access to bona fide topological states.

\section{Frustration, Topology, Ice, and Spin Ice}

\begin{figure}[t!]
\includegraphics[width=.7\columnwidth]{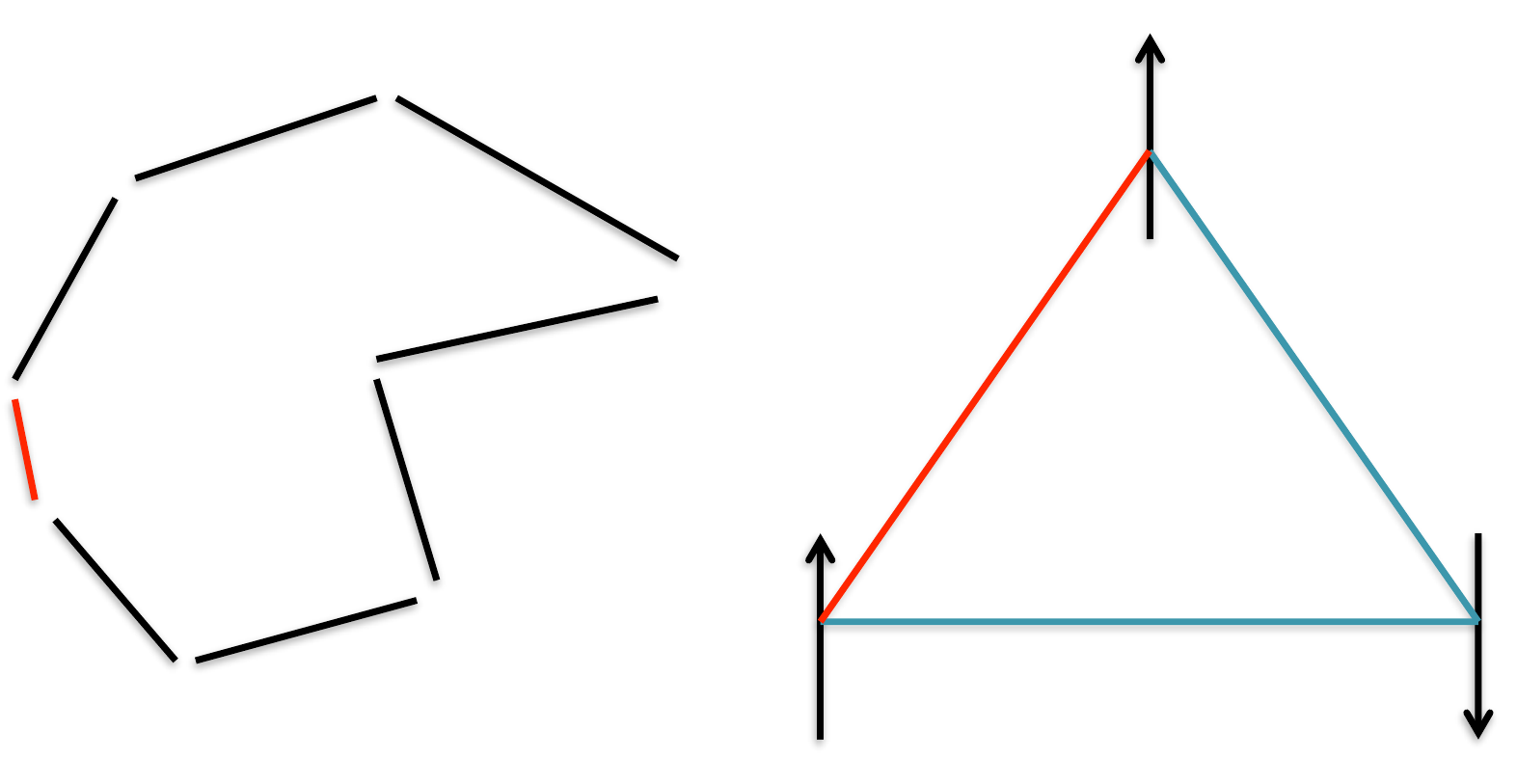}

\begin{center}
\caption{Geometric frustration can be understood schematically as a set of prescriptions that cannot be satisfied simultaneously around certain loops. The red link on the figure on the left represents an ``unhappy link" in a generally frustrated system. More specifically, for an Ising antiferromagnet (right) the loop in question is a loop of interactions among nearest neighbors. On a triangular lattice, triangular loops are frustrated, as one of the three links (red) must be unhappy. }
\label{default}
\end{center}
\end{figure}

The concept of geometric frustration in its broader mathematical form involves a geometric system, a manifold of degrees of freedom and a set of prescriptions on how they should arrange with respect to each other. The system is frustrated if there are loops along which not all these prescriptions can be satisfied (Fig.~1). Clearly the concept is very general and extends beyond Physics. One recognizes topology immediately in the nature of such definition: any homotopy, that is any continuous transformation that does not tear those loops, will lead to a system of the same frustration. 

In Physics, in general (a)  these ``prescriptions'' correspond to  the optimization of  a certain energy, and (b) that energy is usually a pairwise interaction between binary degrees of freedom. 

An early example is  the famous antiferromagnetic Ising model on a triangular lattice~\cite{wannier1950antiferromagnetism}, a system of binary spins interacting antiferromagnetically on a triangular latice  (Fig. 2). There the  interaction among nearest neighbor spins cannot be satisfied simultaneously on a triangular plaquette, leading to a disordered manifold. The disorder is, however, non-trivial, and  its entropy per spin is not merely $s=k_B\ln(2)\simeq 0.6931k_B$, because rules apply, due to frustration: of all the energy links, only one per plaquette is frustrated, and it relieves the frustration on adjacent plaquettes. Indeed its entropy per spin at $T=0$ is $s\simeq0.3383 k_B $, different from zero, in violation of the third law of thermodynamics, and about half of the entropy of a completely random configuration.

However, as most realistic interactions in Physics are   geometric (for instance, the dipolar interaction between magnets is anisotropic) they immediately break the topological structure of frustration in a real system. We will discuss later  how renouncing  the point~(b)  opens the way to great freedom of design in artificial spin ices. 

Perhaps the first  famous occurrence of frustration in the history of Physics pertained to water ice. In the 1930s Giaque and Ashley~\cite{giauque1933molecular,giauque1936entropy} performed a series of carefully conducted calorimetric experiments, deduced the entropy of water ice at very low temperature, and found that it was not zero. The answer to this mystery would be provided by Linus Pauling a few years later~\cite{pauling1935structure}. Ice comes in many crystalline forms, but all imply oxygen atoms residing at the center of tetrahedra, sharing four hydrogen atoms with four nearest neighbor  oxigen atoms (Fig.~3). Two of such hydrogens will be covalently bond, and two will realize an hydrogen bond: two are ``in", two are ``out"  of the tetrahedron. This is the so called ice-rule previously introduced by Bernal and Fowler~\cite{bernal1933theory}. Each tetrahedron has $6$ admissible configurations out of the $2^4=16$ ideally possible, and the collective degeneracy  grows exponentially in the number of tetrahedra $N$ as $W^N$, leading to a non-zero entropy per tetrahedron $s=k_B \ln W$ for this disordered manifold. In what can be considered as one of the most precise and felicitous  back-of-the envelope estimate in the history of statistical mechanics, Pauling counted such degeneracy as $W=3/2$, remarkably close to both the experimental value and to the numerical value ($W=1.50685 \pm 0.00015$~\cite{nagle1966lattice}). 

\begin{figure}[t!]
\includegraphics[width=.4\columnwidth]{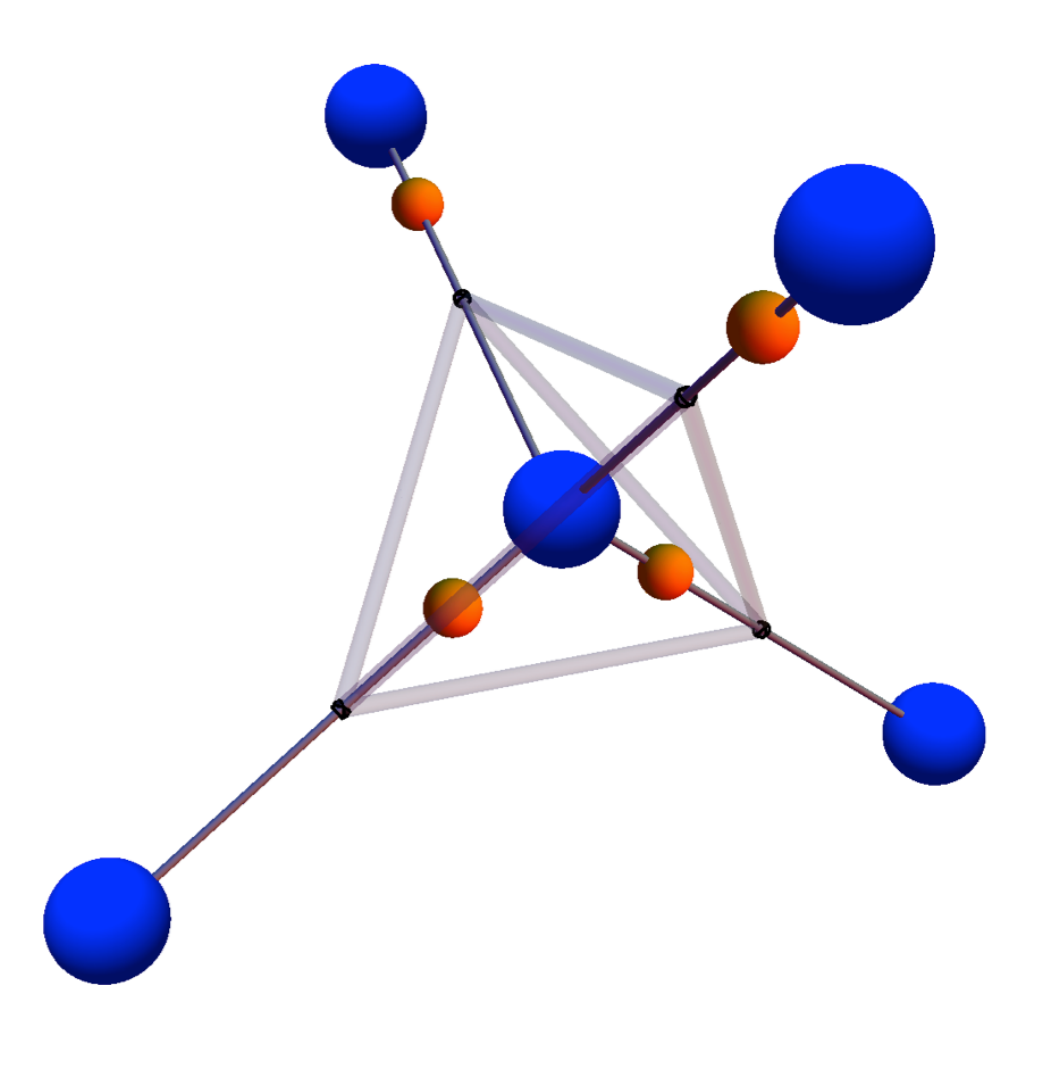}\includegraphics[width=.4\columnwidth]{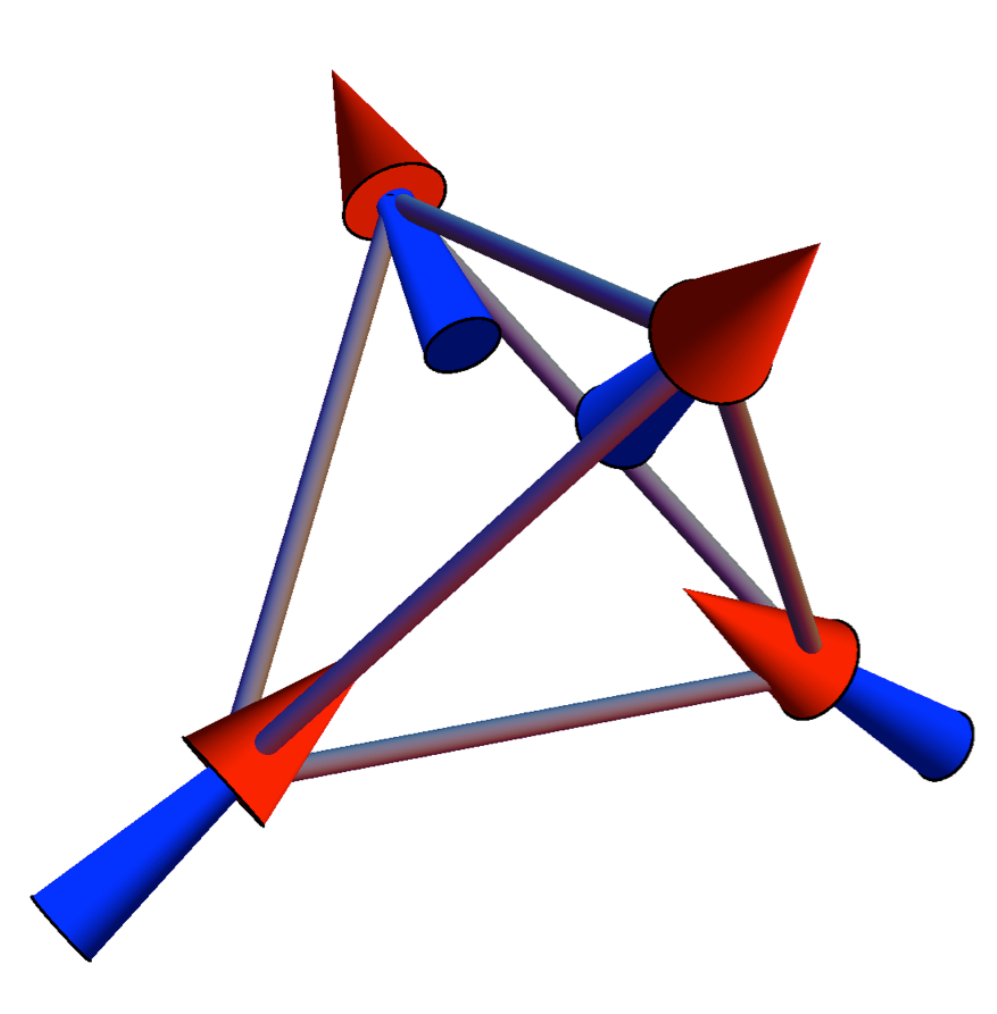}
\begin{center}
\caption{The ice rule. Left: In water ice oxygen atoms sit at the center of tetrahedra, connected to each other by a hydrogen  atom. Two of such protons are close (covalently bonded) to the oxygen at the center, two are further away, close to two of the four neighboring oxygens. Right: One might replace this picture with spins pointing in or out depending on whether the proton is close or far away. Then two spins point in, two point out. This corresponds to the disposition of magnetic moments on pyrochlore spin ices, rare earth titanates whose magnetic ensemble does not order at low temperature, because of frustration, and, much like water ice, provides non-zero low temperature entropy density. (Figures from ref~\cite{Castelnovo2008}).}
\label{default}
\end{center}
\end{figure}

These works pointed to the reality of exotic disordered states in the most common and vital substance on earth. Decades later, they also motivated the introduction by Lieb, Wu, and Rys  of simplified models of mathematical physics, known as vertex models, which could in many cases be solved exactly~\cite{lieb1967residual,lieb1967exact,Wu1969,Baxter1982, rys1963ueber}. Those are two dimensional models of in-plane spins impinging on vertices, where different energies are associated to different vertex-configurations, and whose statistical mechanics is usually solved via transfer matrix methods.

Ice-like systems have then received renewed interest in the 1990s, when unusual behaviors were discovered in the low temperature regime of rare earth titanates such as Ho$_2$Ti$_2$O$_7$ whose magnetic moments exhibit a net ferromagnetic interaction between nearest neighbor spins, yet no ordering at low temperature,  suggesting strong frustration. Similar to protons in water ice,  the magnetic moments of these materials reside on a lattice of corner-sharing tetrahedra, and they are constrained to point either directly toward or away from the center of a tetrahedron (Fig.~3). The resulting ferromagnetic interaction favors a 2-in/2out ice-rule. The similarity noted by Harris et al ~\cite{harris1997geometrical}  was confirmed experimentally  by Ramirez et al.~\cite{Ramirez1999}.

Besides providing  important model systems with novel field-induced phase transitions and unusual forms of glassiness, and an early and practical example of a classical topological state, spin ices harbor a new fractionalization phenomenon in their low energy dynamics: emergent magnetic monopoles~\cite{ryzhkin2005magnetic,Castelnovo2008}.  
\begin{figure}[t!]
\includegraphics[width=.9\columnwidth]{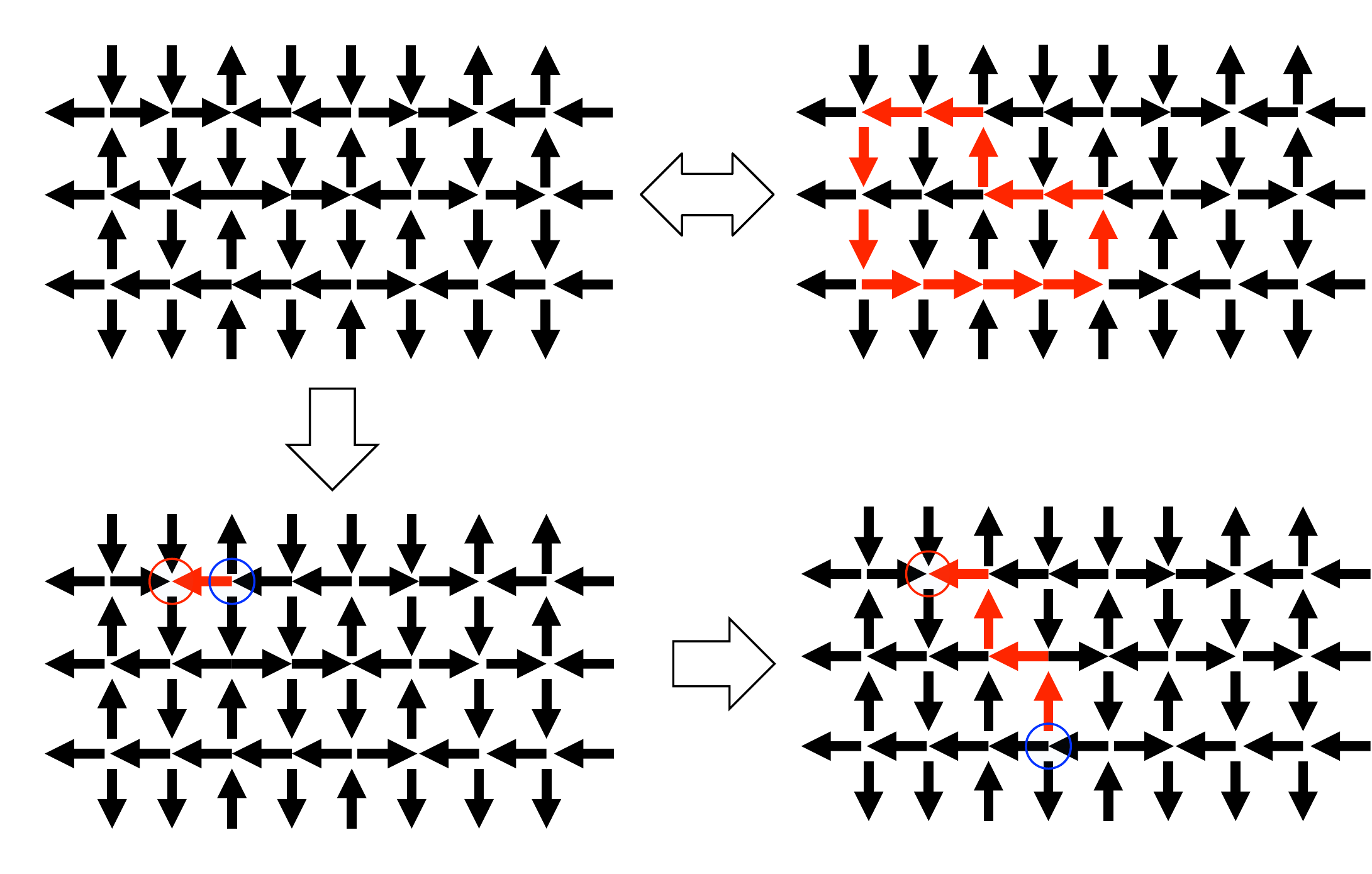}
\caption{The ice manifold (top left), an ensemble of spins obeying the ice rule (in each vertex two arrows point in, two out). One can obtain another realization of the ensemble only by flipping a proper loop of spins. Flipping a single spin creates a couple of magnetic monopoles of opposite charge (positive is red, negative is blue). The monopoles can be separated by further spin flips (creating a ``Dirac string'', shown in red), interact via Coulomb interaction, but are topologically protected, as they can only be created and annihilated in pairs.  }
\label{default}
\end{figure}
To facilitate understanding consider the two-dimensional schematics of Fig.~4, which  represent a disordered ice manifold, an ensemble of spins where all the vertices obey the ice rule. The reader will notice that it is impossible to explore the  manifold by single spin flips, without breaking the ice-rule. Only by flipping proper loops of spins we can obtain a new configuration within the ice-manifold. This is already a hint of the topological nature of the state.

If we flip one spin only, we create two defects (3-in/1-out and vice versa). We can separate those defects by further flips, and we have two deconfined magnetic monopoles, and one can prove through multipole expansion  that their interaction is Coulomb~\cite{Castelnovo2008}. Of course these monopoles are in effect simply the opposite ends of a long, floppy dipole, in red in figure, called the Dirac string; however, owing to the disorder of the manifold, the system is no longer reminiscent of the Dirac string connecting the monopoles. Thus, excitations over the ice manifold can be described by a fractionalization of the spins into individual, separable magnetic charges which interact via a Coulomb law.

In view of the  more complex geometries that we will discuss later, let us generalize the notion of ice manifold and ice rule for a general lattice, or graph, or network~\cite{mahault2017emergent}, whose edges are spins impinging in vertices of various coordination $z$. Then we say that a vertex of coordination $z$ with $n$ spins pointing toward it has {\it topological charge} $q=2n-z$, corresponding to the difference between spins pointing in and out. In general, we call spin ice systems those in which $|q|$ is minimized locally at each vertex (typically, but not necessarily, by nearest neighbor spin-spin interaction). For a lattice of even coordination, such as the square ice or pyrochlore ice introduced before, the ice manifold is characterized by zero charge on each vertex. However, for lattices of odd coordination there cannot be any charge cancellation, and thus in the ice manifold each vertex will have charge $q=\pm1$ in equal fraction, as the total charge of a system of dipoles must always be zero. That is the case of Kagome ice, which we will discuss in the next section.

These magnetic charges are topologically protected monopoles in square or pyrochlore ice: their magnetic charge is indeed also a topological charge and one can see from Fig.~3 that charges can only be created and annihilated in opposite pairs. One could indeed create a single monopole in an open system  but that would simply imply pushing the second one at the boundaries. If we, however, placed the ensemble of Fig.~3 on a torus, thus without boundaries, then clearly there would not be a net monopole charge in the bulk. Even in an open system, the total net charge will be proportional to the flux of the magnetic moment through the boundaries, and as the latter is bound by the net magnetization of the spins, one finds that the density of net charge must scale with the reciprocal length of the system.  These two-dimensional considerations extend to the three-dimensional spin ice.

In more theoretical terms, the topological state of spin ice is an example of a so called Coulomb phase~\cite{henley2010coulomb},  a phase described not by an order parameter or a symmetry breaking, such as the ordered phases that fall within the Landau paradigm, but rather by a solenoidal emergent field (the coarse grained magnetization $\vec M$). One can say that the ice rule and its charge cancellation corresponds to a divergence-free condition $\vec \nabla \cdot \vec M=0$ on each vertex. Then, a monopole in $\vec x_0$ is a source of divergence $\vec \nabla \cdot \vec M=q_m\delta(\vec x -\vec x_0)$ of charge $q_m=Q q$ where $Q=M/L$ (the magnetic moment  $M$ of the spin divided by its length $L$), and q is the topological charge of the vertex defined above. This is   considered as an example of {\it classical topological order}~\cite{castelnovo2012spin}. 

Indeed, while quantum topological order  has provided a valuable framework to conceptualize disordered states of spin liquids that escape a Landau symmetry breaking paradigm and cannot be obviously characterized by local correlations~\cite{wen1989vacuum,wen2002quantum}, the importance of topological states had been recognized even earlier in classical physics~\cite{chaikin2000principles}: in the  theory of dislocations~\cite{volterra1907equilibre}, liquid crystals~\cite{kurik1988defects}, or topological transitions~\cite{kosterlitz1973ordering}. Recently, whether in direct analogy with quantum physics~\cite{castelnovo2007topological}, in purely abstract terms~\cite{henley2011classical,lamberty2013classical}, or motivated by real systems such as pyrochlore spin ices~\cite{castelnovo2012spin,jaubert2013topological,henley2010coulomb}, a consistent notion of  {\it classical} topological order in {\it discrete} systems has  been proposed, to conceptualize  (i) a degenerate, locally disordered  manifold (ii) described  by a topologically non-trivial, emergent field  (iii) whose  topological defects (in spin ice, magnetic monopoles~\cite{ryzhkin2005magnetic,Castelnovo2008})  coincide with excitations above the manifold. 

Topological protection implies  that  states {\it within} the  manifold can be linked only via collective changes of entire loops of a discrete degree of freedom. Thus any realistic low-energy dynamics  happens necessarily  {\it above the manifold}, through  creation, motion, and annihilation of pairs of protected topological excitations. Typically, their constrained and discrete kinetics  leads to  ergodicity breaking, fractionalization and thus various forms of glassy behaviors~\cite{henley2011classical,castelnovo2010thermal}.  We will see later that such order can also be found in novel, non-trivial geometries of artificial spin ice characterized by vertex-frustration, such as Shakti spin ice.

\section{Simple Artificial Spin Ices}

After introducing the main concepts, we warm up to the field of artificial spin ice by summarizing briefly the early work on classical geometries, based on the square and honeycomb lattices. Further, more general details about fabrication anc characterization will be discussed in the context of these early realizations. 

\subsection{Kagome Spin Ice}

Even before artificial spin ice realizations~\cite{tanaka2006magnetic,qi2008direct,Nisoli2010} (Fig.~5)
honeycomb structures have been extensively studied theoretically as they describe the two-dimensional behavior of the three dimensional spin ice pyrochlores under a magnetic field aligned along a particular crystalline axis. A honeycomb ice is often called the Kagome spin ice, as the spins reside on the edges of a honeycomb lattice, which is a Kagome lattice, the honeycomb dual lattice. In the context of artificial spin ice, Kagome represented the only simple geometry with a degenerate ice manifold. Indeed, as we will see in the next section, square ice has a frustrated yet perfectly ordered, antiferromagnetic ground state.

\begin{figure}[t!]
\includegraphics[width=.5\columnwidth]{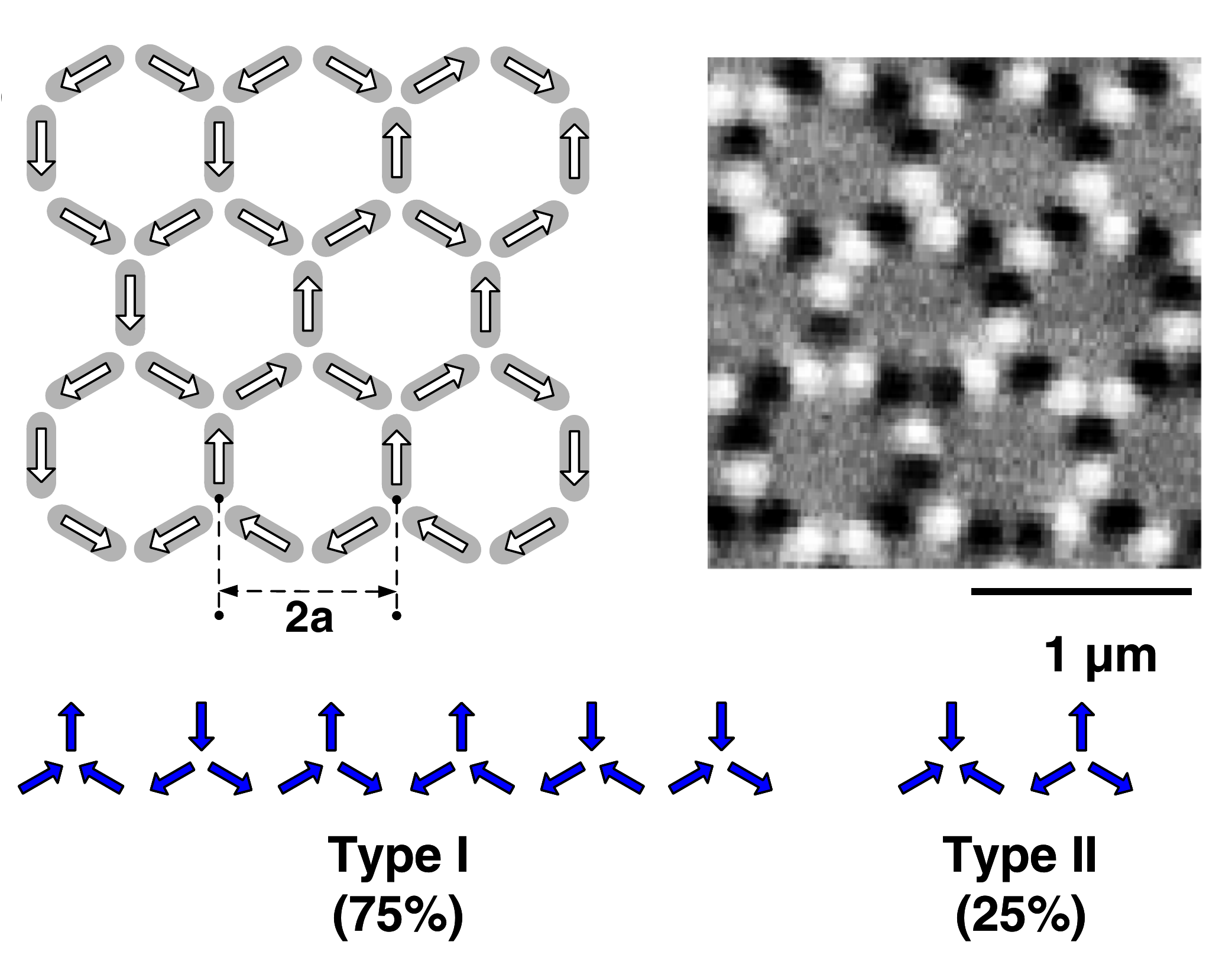}\includegraphics[width=.33\columnwidth]{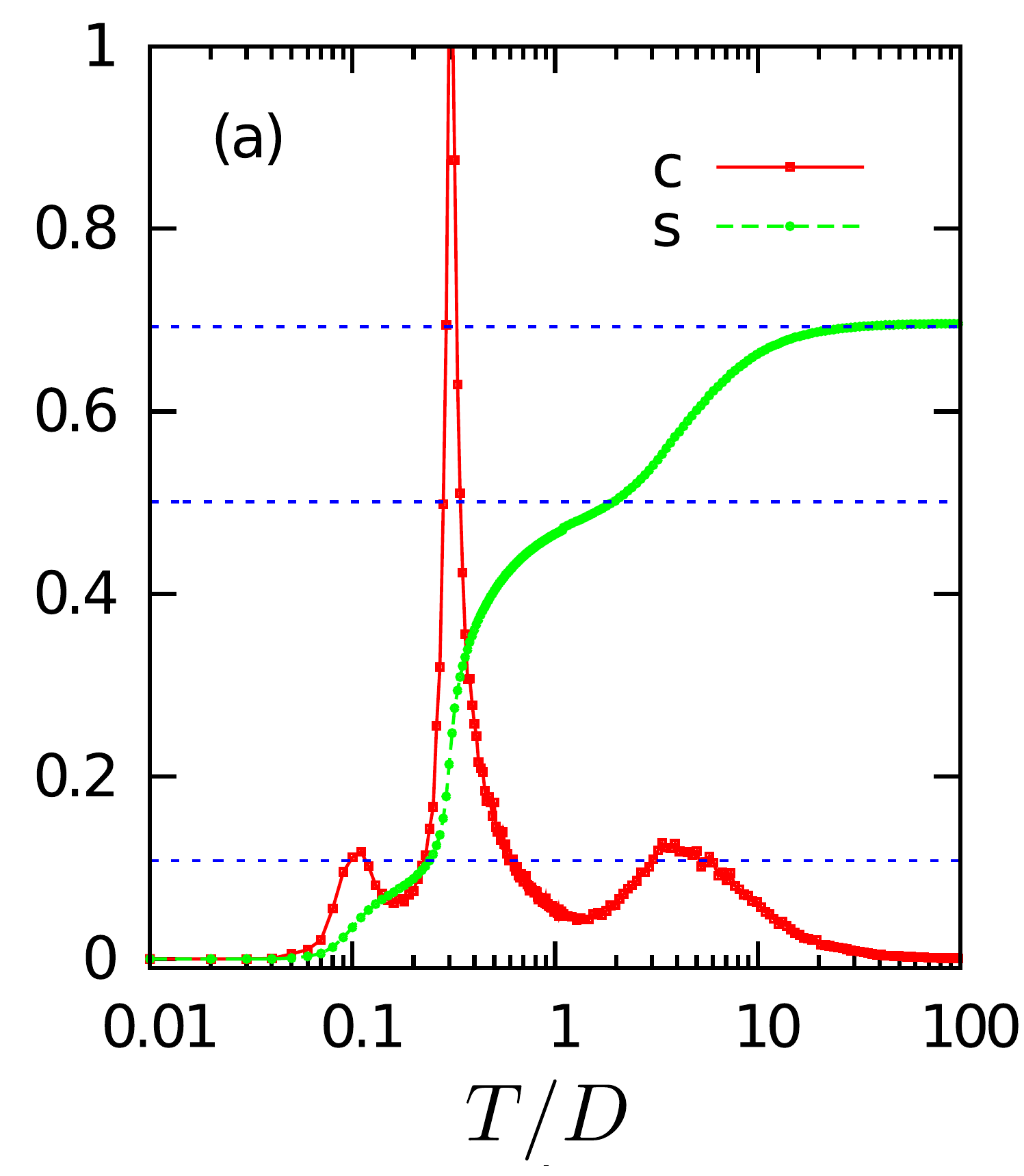}
\includegraphics[width=.8\columnwidth]{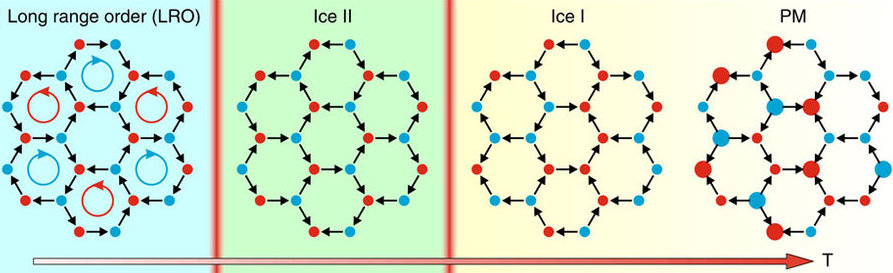}
\caption{Top, from left to right: Schematics and MFM image of the hexagonal arrays with the 8 vertices of the honeycomb/Kagome artificial spin ice. White arrows show the vertex Ice I state, and the percentages indicate the vertex multiplicity. Type-I vertices have lower energy than Type-II and correspond to the generalized ice rule. Temperature dependence (top right) of the specific heat $c$ and entropy per spin $s$  of the Kagome spin ice obtained by ref~\cite{Chern2013}. The dashed lines show values of entropy per spin $s=0.693$ (Ising paramagnet), 0.501 (Ice I), and 0.108 (charge-ordered spin ice, or Ice II). Bottom: the four phases of Kagome ice ordered by increasing temperature. Figures are adapted from ref~\cite{Nisoli2010,Chern2013,anghinolfi2015thermodynamic}.}
\label{default}
\end{figure}

Before proceeding with Kagome ice, some more general details on  artificial spin ice materials are in order, starting with the energetics involved. In general, magnetic, elongated nano-islands can be described as nano--spins, binary degrees of freedom describing their magnetization along their principal axis. This is, however, already an approximation of the magnetic texture of the nano-structure: indeed both direct characterization and micromagnetic simulations show potentially significant relaxation of the magnetization field at the tips of the islands, due to the local field of the surrounding islands. 

A further approximation, which seems to work surprisingly well, implies describing the inter-island interaction via a vertex model. There we assign energies to the various vertex configurations as in Fig.~5. The nano-islands being magnetic dipoles, one expects this approach to eventually break down. It does indeed in enticing ways, revealing inner, low entropy phases within the ice manifold.  

The equilibrium phases of the system have been investigated numerically~\cite{Moller2009,Chern2011} via Metropolis Monte-Carlo simulations with full dipolar interaction. Figure 5 shows  that at high temperature the system is paramagnetic. As temperature is reduced, we cross over toward a disordered ice-manifold, called Ice I, where  each vertex has  charge  $\pm1$. This is as much as a vertex model approximation would explain, as the ice-rule minimizes the energy of the vertices.

However, at lower temperature, we see a transition toward charge ordering: the disordered plasma of magnetic charges residing on the vertices orders within an ionic crystal. The transition is of the Ising class and is due to the Coulomb interaction among magnetic charges. It can be replicated within a vertex-model approximation only by adding further interaction via Coulomb coupling between the charges of the vertices. Note that such state, called Ice II, while being charge ordered, is still disordered in the spin structure, as there are an exponentially growing (in the number of spins) number of possible spin configurations that correspond to the charge ordered state. Finally, further lowering the temperature, another transition leads to an ordered state, where order is brought in by the long range effects of the dipolar interaction. 

\begin{figure}[t!]
\includegraphics[width=1\columnwidth]{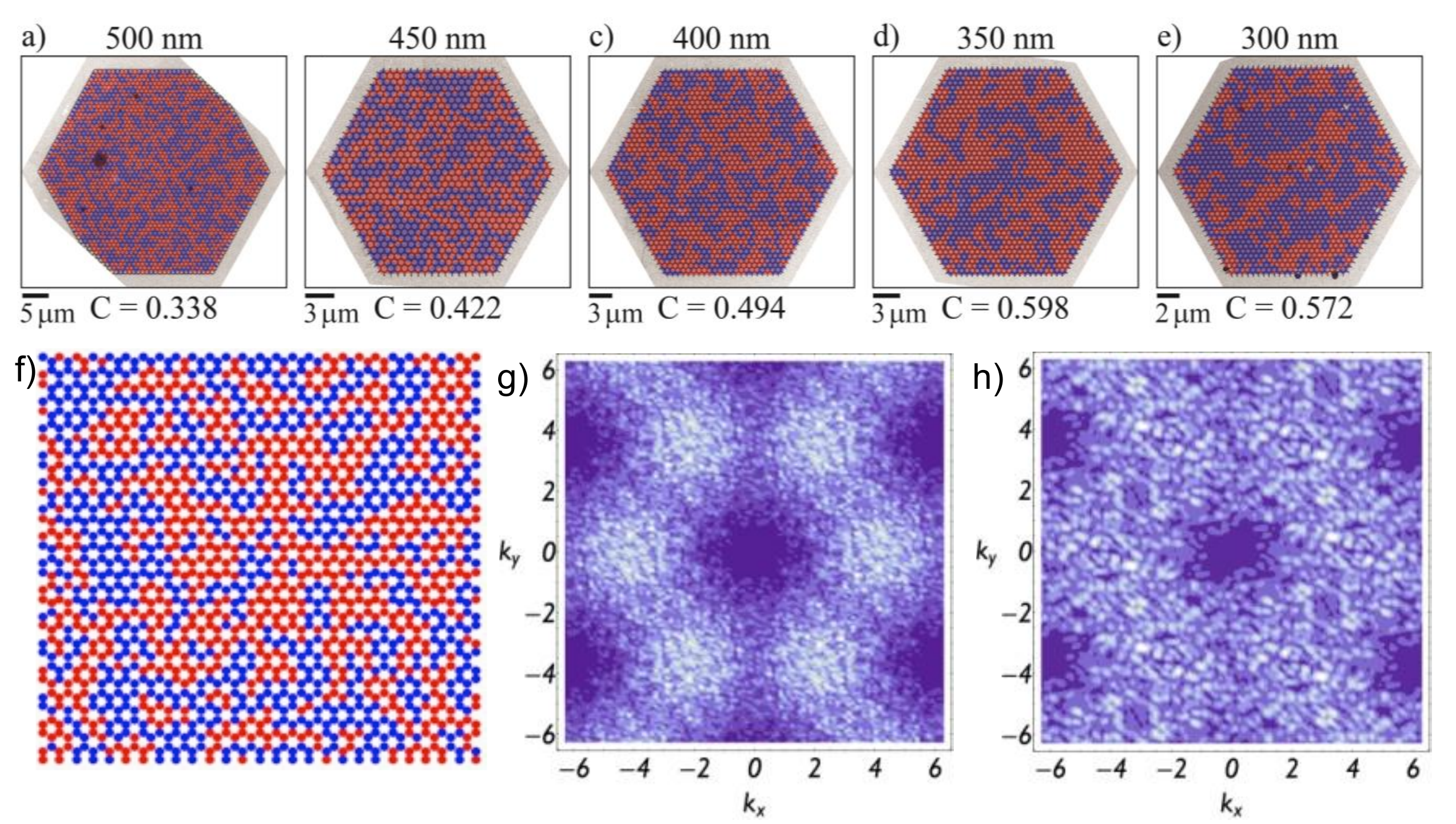}

\caption{Magnetic charge ordering in Kagome ice. (a)-(e) Charge domain maps obtained via Lorenz TEM relative to the annealing of Fe-Pd alloy artificial Kagome ices of different edge length (from 500 nm to 300 nm)  show increasing size of the ionic crystallites of charges as the lattice constant decreases, and thus the mutual interaction among magnetic charges increases. $C$ is the charge-charge correlation parameter ($C=1$ for a fully charge-ordered state). Images adapted from ref~\cite{drisko2015fepd}. (f) Charge map obtained via MFM after annealing of permalloy Kagome ice of lattice constant 260 nm showing incipient domains of charge-ordering and (g) its static structure factor showing incipient peaks corresponding to crystalline order. (h) Static structure factor for lattice constant 490 nm, showing no incipient peaks.}
\label{default}
\end{figure}

These states were variously investigated experimentally. Ice I proved easy to reach. Indeed, even non-thermal methods were able to reach it~\cite{tanaka2006magnetic,qi2008direct,Nisoli2010}. Those methods pertains to thicker islands that are thus not superparamagnetic at room temperature (that is, do not flip their magnetization under thermal fluctuations). These islands are therefore coercive enough that MFM would provide a non-destructive characterization at room temperature. The AC demagnetization~\cite{Ke2008} of such samples is sufficient to reach the ice manifold, a fact which already points to its lack of topological protection. 

The facility with which such state could be reached is telling. Indeed, while  magnetic charges are topologically protected in pyrochlore ices~\cite{castelnovo2012spin}, as we saw above, they are not bona fide topological numbers in the ice manifold of Kagome ice. There, vertices of odd coordination can gain and lose charge freely from the surrounding, disordered, and overall neutral plasma of magnetic charges. Consequently, the ice-manifold can be explored {\it from within} by consecutive single-spin flips, without any need for collective loop-moves such as those shown in Fig.~4. 

Instead, the charge-ordered state, or Ice II, cannot be explored by individual spin flips. A glimpse of the Ice II phase shown in Fig.~5 should convince that any spin flip within the manifold will lead out of the manifold, as it will locally destroy the charge order.

The Ice II state has been explored via thermal methods capable of providing a bona fide thermal spin ensemble. These are of three kinds: annealing at higher than room temperature followed by characterization at room temperature~\cite{Porro2013,Zhang2013,drisko2015fepd}; thermalization with real-time, real-space characterization~\cite{kapaklis2014thermal,Farhan2013, gilbert2016emergent}; and thermalization without real-space characterization~\cite{Kapaklis2012,anghinolfi2015thermodynamic}. In the first, the material is not superparamagnetic at room temperature, but it is heated slightly above the Curie temperature of the nano-islands (which can vary, depending on the size and chemical composition of the nano-structure, from about 600 Celsius for permalloy down to about 100 Celsius for Fe-Pd alloys) and then annealed down into a frozen state, usually characterized via MFM. In the second method the nano-islands are chosen to be thin enough (usually  thickness of 2-3 nm) to be superparamagnetic at room temperature or below, and thus need to be characterized via PEEM, at a proper beam source. In the third, various averaged quantities are extracted, such as the average flip rate of spins, through muon spectroscopy~\cite{anghinolfi2015thermodynamic}---while spin noise spectroscopy~\cite{0034-4885-79-10-106501,crooker2004spectroscopy} should also be a viable method.

Figure~6 shows the results of thermal annealing on artificial hexagonal ice made of permalloy~\cite{Zhang2013}, which demonstrate formation of crystallites of magnetic charges, due to the Coulomb interaction between the charges themselves. Control over the size of those ionic crystallites has also been obtained by employing an alloy of iron and palladium, rather than permalloy, which has a much lower Curie temperature~\cite{drisko2015fepd}. However, nobody has yet  reported any direct evidence of complete long range charge order in such a material, nor of the zero entropy phase of spin order (Fig.~5).

Indirect indications that such low entropy phases within the Kagome ice manifold---or  at least some kind of phases---can be reached were obtained via muon spectroscopy studies, involving islands that were too small and therefore too active to be imaged directly, but whose rate of magnetic flipping could be deduced from the relaxation time of muons implanted on a gold cap over the two-dimensional array. There, the critical slowing down of the spins was measured and found to correspond to that of the numerically predicted transitions, where parameters for the numerical simulations were taken from the material~\cite{anghinolfi2015thermodynamic}. These results, the first to probe deep inside the ice manifold of Kagome lattice,  represent a strong corroboration of the existence of a complex phase diagram, most likely the theoretical predicted one. However, we should not forget that topological or ordered states can be  hard to reach via spin dynamics of the Glauber kind~\cite{glauber1963time}, and that indeed the actual spin dynamics might be rather more complex than a Glauber model. Indeed these ``spins'' are  nanoscopic objects with their own magnetic reversal dynamics. Such specificity might bias certain kinetic pathways, leading to non-equilibration or ergodicity breaking even in Ising models that are not susceptible to these phenomena: thus the phases whose critical slowing down was experimentally revealed could be only reminiscent of the one predicted at equilibrium, which of course adds to their potential interest. 

\subsection{Square Ice}

With the exception of the work of Tanaka {\it et al}~\cite{tanaka2006magnetic}, early works concentrated on the square geometry (Fig.~1)~\cite{Wang2006, Nisoli2007, Nisoli2010}. Square ice also represented the benchmark on which to test demagnetization and annealing methods which lead to experimental protocols for thermal ensembles~\cite{Morgan2010,Kapaklis2012, Porro2013}. 

\begin{figure}[t!]
\includegraphics[width=.67\columnwidth]{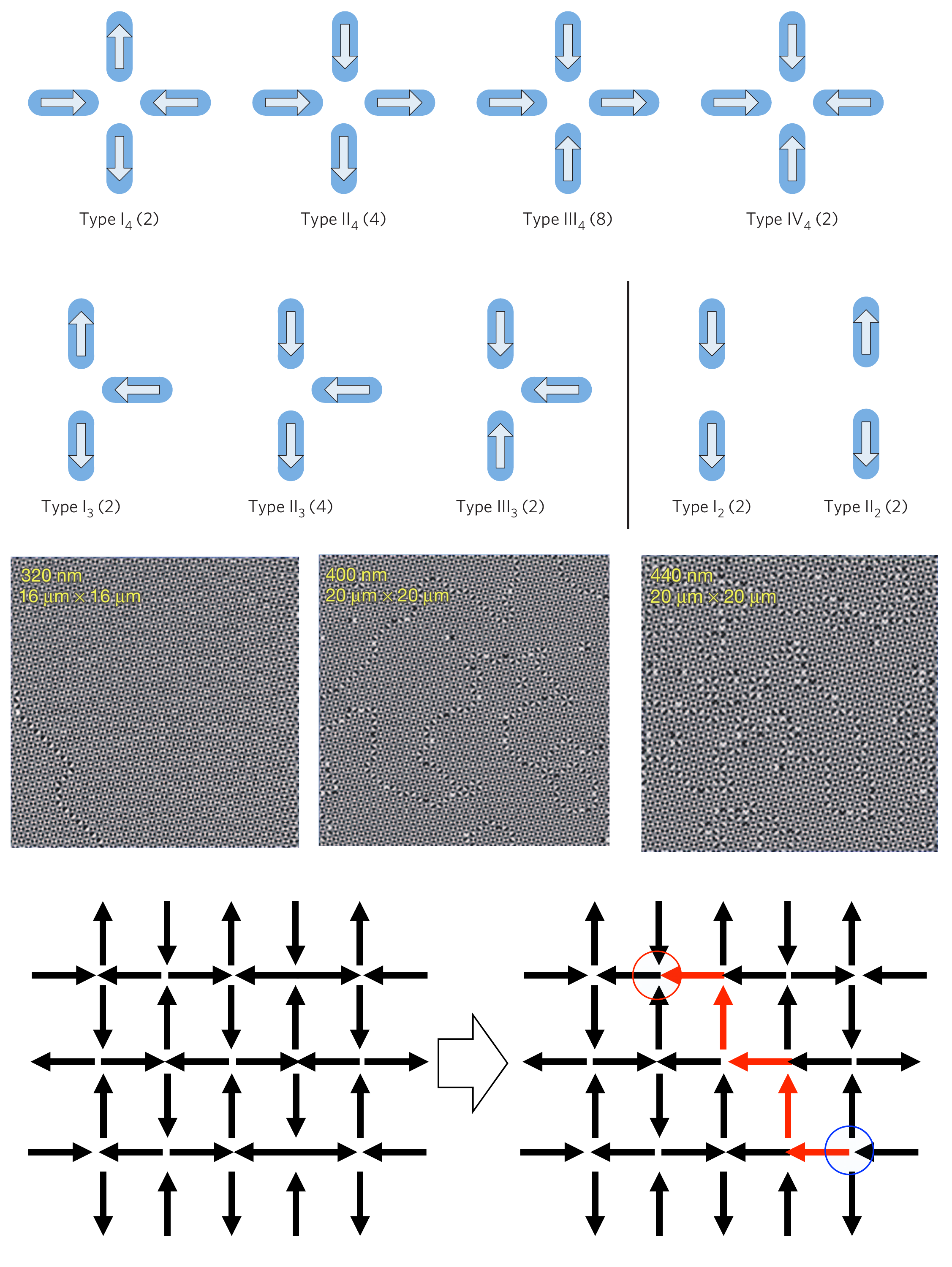}
\caption{Top: Vertex-configurations for $90\degree$ angles of coordination $z=4,3,2$  (degeneracy in brackets) listed in order of increasing energy. Middle: MFM images of thermally annealed square ice at different lattice constants showing an ordered domain crossed by a Dirac string (for the specimen at 320 nm) and a multi-domain ensemble separated by domain walls of monopoles and diract strings (at 400 nn and 440 nm); note also the frozen in monopole pairs (figures adapted from ref~\cite{Zhang2013}). Bottom: the lowest energy state of square ice as an antiferromagnetic tiling of Type-I$_4$ vertices; creating and separating a monopole pair entails a Dirac string (red) of Type-II$_4$ vertices, that are energetically more costly than the Type-I$_4$, leading to the linear confinement of the pair.}
\label{default}
\end{figure}

It is important to understand that square artificial spin ice does not resemble the square ice of Lieb~\cite{lieb1967residual}, or the degenerate square ice described in Fig.~ 4, firstly because it admits topological defects in the form of magnetic monopoles absent in the six-vertex model, but most importantly because it is not degenerate. In this sense it shares similarities with the Rys F-model~\cite{rys1963ueber} but those should not be overstated, as the (physically unnatural)  absence of monopoles in the latter leads to an infinitely continuous  transition to antiferromagnetic ordering~\cite{lieb1967exact}, whereas in the former the transition is of second order.

Figure~7 shows the energetic hierarchy of vertices with $90\degree$ angles (including those of coordination $z=3,2$ to be discussed later). Because of the anisotropy of the dipolar interaction, nearest neighbor perpendicular islands interact more strongly  than collinear ones, leading to lifting of degeneracy within the ice manifold. The system, if modeled at the vertex level, can be described as a $J_1, J_2$ antiferromagnetic Ising model on a square lattice, with a transition to antiferromagnetic ordering, which indeed has been obtained experimentally via thermal annealing, as shown in Fig.~7. 

Within the ordered state of  square ice potentially interesting transitions have been proposed~\cite{mol2010conditions, Nascimento2012,Mol2009}: when the system is not degenerate, creating and separating a couple of monopoles requires energy proportional to the number of Type-II in the Dirac string (see Fig.~7). Much like quarks or Nambu monopoles~\cite{nambu1974strings} these pairs are linearly confined, and the tensile strength of their  Dirac string drives the ordering as the temperature is reduced. However, one can imagine that a topological transition corresponding to monopole deconfinement might take place under proper conditions when the energy of the Dirac string is offset by its fluctuating entropy. 

Square ice, however, can be made properly degenerate, which means described by a spin ensemble such as the one of Fig.~4, revealing an emergent topological Coulomb phase. One way is to raise the vertical islands with respect to the horizontal ones ~\cite{Moller2009}, a method that has recently been pursued experimentally~\cite{perrin2016extensive} demonstrating a degenerate manifold whose static structure factor coincides with the numerically computed one for a six-vertex model, thus providing the first artificial realization of a two dimensional Coulomb phase. We have also proposed  to iterate such design on the axis perpendicular to the array, and we have designed layered structures that are geometrically different but topologically equivalent to three dimensional spin ice pyrochlores~\cite{chern2014realizing}. Those have not found  realization yet. The only three dimensional realization of artificial spin ice was obtained by filling the voids of an artificial opal film with Cobalt~\cite{mistonov2013three,mistonov2015ice}, a promising approach to bring to room temperature some of the features of spin ice pyrochlores. Of course, as always with three dimensional realizations, the challenge there lies not only in nano-fabrication, but also in characterization, as real-space methods are generally surface methods. 
 
Finally, another way to produce a Coulomb phase in square ice has been presented recently, and involves ``rectangular ice'' where vertical and horizontal islands differ in length, and degeneracy is obtained for a proper critical value of their ratio~\cite{Nascimento2012,ribeiro2017realization}.

\section{Exotic States Through Vertex-Frustration}

Until 2014, the only degenerate artificial spin ice was Kagome. As both nano--fabrication and characterization protocols evolved, it became clear that the initial inspiration of the entire project---to design exotic behaviors in the geometry of interacting, binary  degrees of freedom---could become viable, if not for one problem: in real systems, the frustration of the pairwise interaction is wedded to the geometry. What this means is explained in Fig.~8 where brickwork spin ice and Kagome spin ice are shown to lead to completely different ground states, one disordered, the other ordered, despite the two geometries being  topologically equivalent. Indeed, the dipolar interaction is not topologically invariant, but instead depends very much on the mutual arrangements of the dipoles.

To overcome this limitation and gain freedom in the design of new  materials  capable of  various states and unusual behaviors, the first step is to decouple frustration from geometry. As the pairwise interaction is anisotropic, something else will have to be frustrated. A possibility is the vertex itself. 

\begin{figure}[t!]
\includegraphics[width=.5\columnwidth]{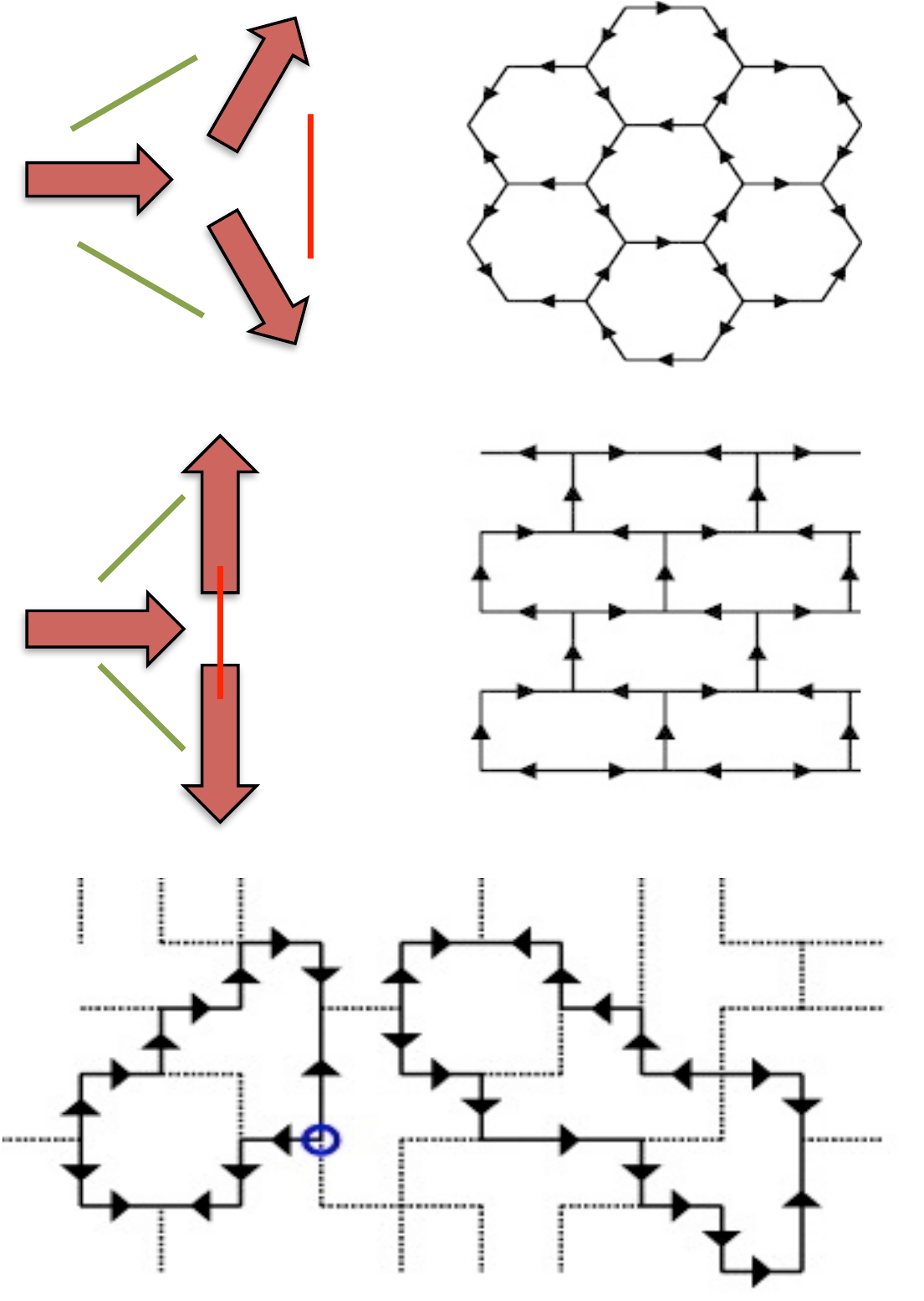}
\caption{Geometry vs. Topology. Topologically equivalent geometries leads to completely different spin ensembles, due to the anisotropy of the dipolar interaction. The honeycomb spin ice (top right) is topologically equivalent to the ladder spin ice (middle, right) yet the nearest neighbor interactions  lead to an ordered ground state in the latter (see also Fig.~7 for the energetic hierarchy of the vertices) and a disordered manifold in the former. Pairwise interactions are frustrated in both systems, however in the honeycomb lattice all the spins interacting in the vertex have the same mutual angle (top left) and thus any of the three interactions can be frustrated, whereas it is energetically favorable to frustrate the interaction between parallel spins in the ladder lattice (middle left). At the bottom is an example of vertex-frustration, where the allocation of vertices of lowest energy is frustrated, leading to ``unhappy vertices'' (blue circles) on certain loops, instead of unhappy energy links (red lines above). }
\label{default}
\end{figure}

Consider a geometry made of $90\degree$ vertices of coordination $z=4,3,2$ (Fig.~7). Each vertex has a unique configuration of minimal energy (up to a flip of all the spins). Imagine now arranging them in such a way that, however, not all vertices can be assigned to the lowest energy configuration~\cite{Morrison2013}. This will lead to ``unhappy vertices'' (UV), that is, topologically protected local excitations (Fig.~8). In proper geometries, the degeneracy of the allocation of such vertices grows exponentially with the size of the system, leading to a degenerate low energy manifold~\cite{Morrison2013, stamps2014artificial}. 

Crucial here is that within this manifold the system is usually captured by an emergent description that considers the allocation of these protected local excitations, rather than by its spin ensemble. As a consequence, other emergent properties appear that are in general not obvious nor indeed apparent in the local spin structure.

While vertex models~\cite{Baxter1982} were introduced to describe frustrated systems, they  were themselves not frustrated. They simply subsumed the degeneracy of a frustrated system within a degenerate energetics. Vertex-frustrated geometries can thus be considered the first frustrated vertex models. Vertex-frustration is of course a nearest-neighbor level concept, although it induces topological states that are collective.  However, the real materials being made of dipoles, other phases are  present within their vertex-frustrated low energy manifold, much like inner phases are present in the diagram of Kagome above. Let us now see how this comes about in three such geometries: Shakti, Tetris, and Santa Fe. 

\section{Emergent Ice Rule, Charge Screening, and Topological Protection: Shakti Ice}

Consider the Shakti geometry in Fig.~9~\cite{Chern2013}. Each minimal, rectangular loop of Shakti is frustrated. What it  means is that  it  must be affected  by an odd number of unhappy vertices~\cite{Morrison2013, Chern2013}. Because each unhappy vertex aways affects two nearby loops  and costs energy, the lowest energy configuration is realized when nearby loops are dimerized by a single unhappy vertex~\cite{Morrison2013}.  If one considers the geometry, one finds (Fig.~9-c) that each plaquette made by two rectangular loops will host two unhappy vertices in 4 possible locations,   much like the ice rule in water ice prescribes that 2 hydrogenatoms are  within the tetrahedron containing each oxygen atom (Fig.~3), in 2 of the 4 possible allocations. In both cases the same ice-rule applies, but here in emergent form:  not in terms of the original spins, but in terms of allocation of unhappy vertices. Thus,  the lowest energy manifold, at the nearest neighbor vertex description employed here, corresponds then to an emergent six-vertex model. This has been shown experimentally~\cite{gilbert2014emergent}.

\begin{figure}[t!]
\includegraphics[width=.7\columnwidth]{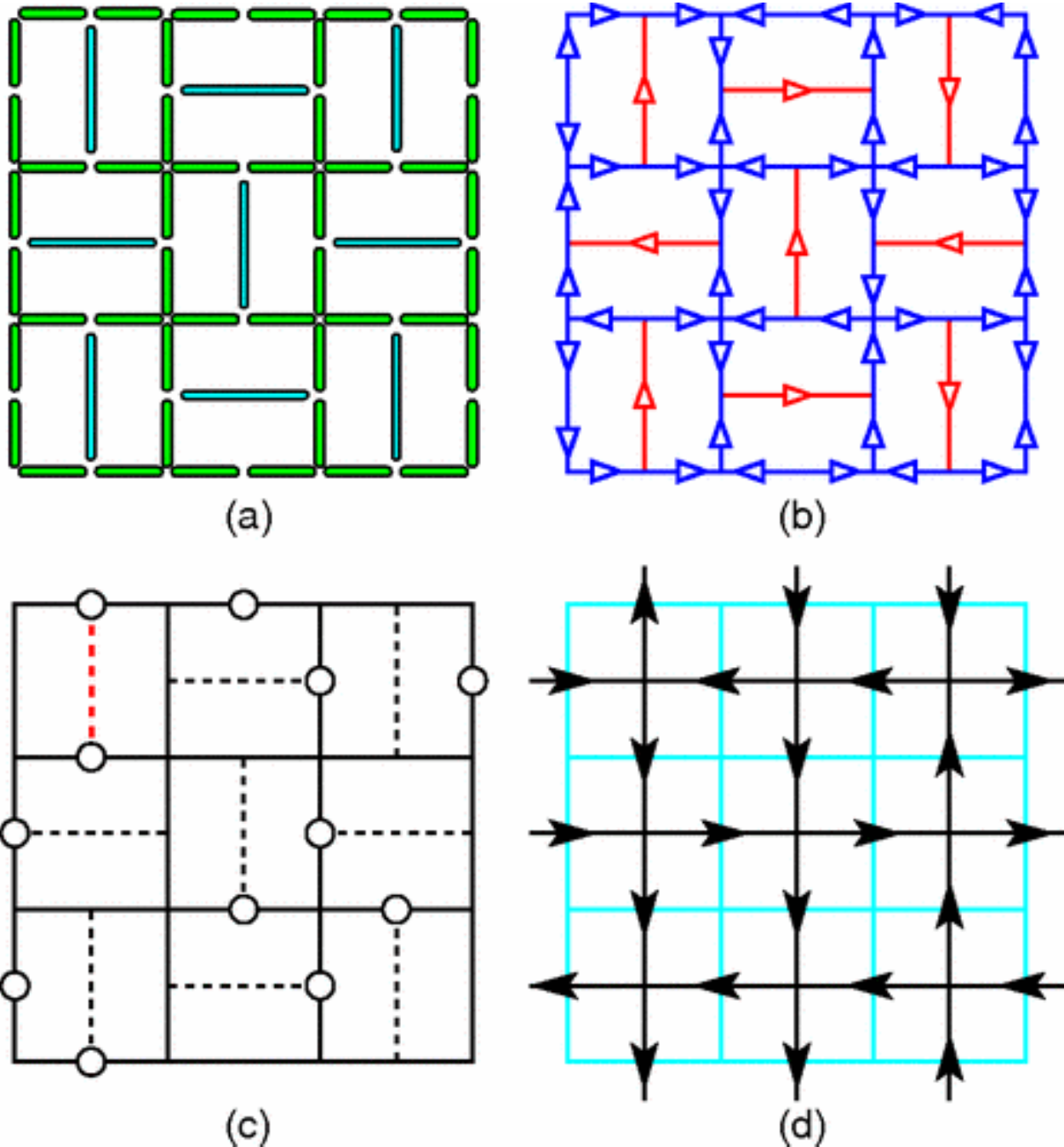}
\caption{Theory of Shakti spin ice. The structure of the system (a) is such that its  lowest energy spin ensemble (b) is disordered. A look at the spin structure in (b) does not seem particularly insightful. However, if we translate that spin map unto the allocation of locally excited vertices, denoted by circles in (c) we then see that each plaquette will host two and only two unhappy vertices in four possible positions. This is equivalent to a six vertex model (d) where  pseudo-spins are assigned to each plaquettes and point toward (away from) the unhappy vertices in plaquette of vertical (horizontal) long island.  Figures adapted from ref~\cite{Chern2013}.}
\label{default}
\end{figure}

\begin{figure}[t!]
\includegraphics[width=.7\columnwidth]{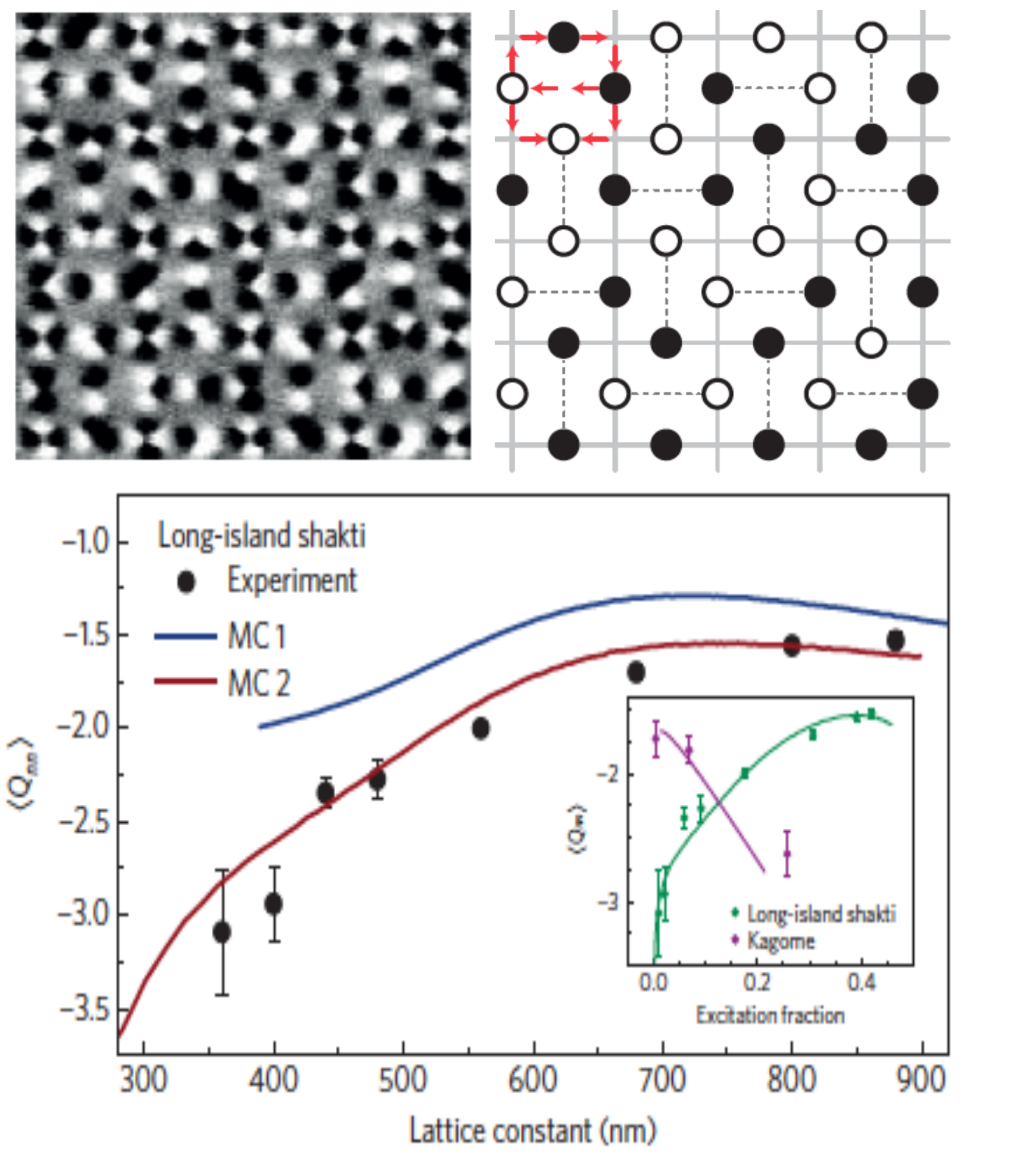}
\caption{Realizations of Shakti  Ice. On the top left, an MFM image of Shakti ice after annealing. On the top right, the experimental data  of the left part is translated in terms of  allocation of the unhappy vertices, on plaquettes (black dots). One can see how an emergent ice rule describes the system, as each plaquette can have only two of four slots for unhappy vertices occupied. Bottom: screening of monopoles from magnetic charges $\langle Q_{nn}\rangle$ denotes the average magnetic charge surrounding a magnetic monopole on a $z=4$ vertex, at the nearest neighbor level. Figures adapted from ref~\cite{gilbert2014emergent}.  }
\label{default}
\end{figure}

Nonetheless, as we had cautioned before, this nearest neighbor description defines the ice-manifold, within which intervene other non-trivial phenomena, due to the long range nature of the interaction. A particularly interesting one regards the screening of magnetic charge. Shakti has multiple coordination, therefore while in its low energy state all the vertices of coordination $z=4$ are in the ice rule, they are surrounded by vertices of coordination $z=3$ which always have a magnetic charge $\pm1$ (in natural units, previously defined), and are disordered. When a vertex  of coordination $z=4$ hosts a magnetic monopole, the overall neutral plasma of charge around it rearranges to screen it, as shown in Fig.~10~\cite{gilbert2014emergent}.

It is important to understand that magnetic monopoles are {\it not} proper topological charges for Shakti, as they are not protected. Each $z=4$ vertex being surrounded by a sea of charges, it can gain or lose charge to and from it. However, the Shakti state is a bona fide topological phase, which means that some other topological charge should be identified in it. 

That the manifold has topological protection can be immediately suspected by noting that a single spin flip takes out of the manifold, and  only a proper loop of collective spin flips realizes change {\it within} the manifold. This can be understood easily from Fig.~9, as all spins impinging in a $z=3$ vertex also impinge into $z=4,2$ vertices, which are in their lowest energy in the manifold. Flipping such spin will thus necessarily cause excitations. 

To identify the topological structure, we go back to the properties of the low energy state. We saw that because each UV affects two nearby plaquettes (Fig. 9) and costs energy, the lowest energy configuration is realized when nearby plaquettes are ``dimerized'' by a single UV~\cite{Morrison2013}. The ice manifold of Shakti is thus described by a dimer cover model on the lattice connecting the rectangular plaquettes, which is topologically equivalent to a square lattice (Fig.~11) (from now on called ``dimer lattice''), and which can be solved exactly~\cite{kasteleyn1961statistics}.

The following is then  standard: a discrete, emergent vector field $\vec E$ can be introduced, perpendicular to each edge, of length 1 (o 3) if the edge is unoccupied (or occupied) by a dimer, and direction entering (exiting) a gray square of Fig.~11 from top or bottom, and exiting (entering) it from the sides.
The ``line integral'' $\int_{\gamma} \vec{E} \cdot d\vec{l}$ for such a discrete vector field along a directed line $\gamma$ crossing the edges is the sum of the vectors along the line with sign taken along  the line's direction. For a complete cover the emergent field is irrotational ($\oint_{\gamma}\vec{E} \cdot d\vec{l}=0$) leading to the definition of a ``height function''~\cite{henley2011classical} $h$ such that $\vec E= \vec \nabla h$ and  thus demonstrating the topological state. 

\begin{figure}[t!]
\includegraphics[width=1\columnwidth]{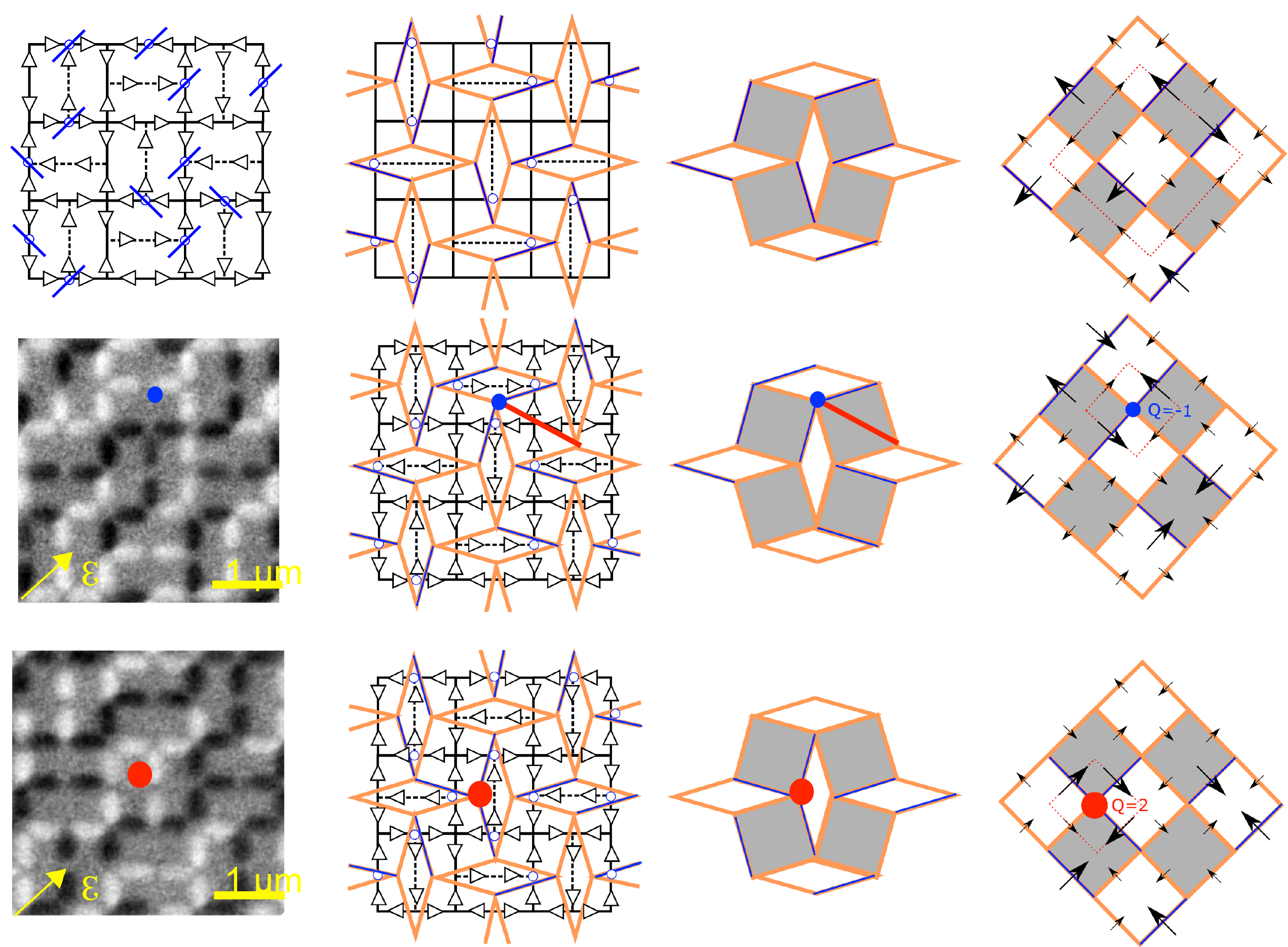}
\caption{Top: Shakti manifold as a dimer cover model. From left to right: Disordered spin ensemble for the ground state of Shakti ice manifold. The manifold is completely described by the allocation of the UVs (circles) which affect two nearby rectangular plaquettes (connected by the blue segments). Thus, an unhappy vertex is a dimer (blue segments) connecting frustrated plaquettes, and the ground state is a complete dimer-cover model on the (Ochre color) lattice with nodes in the center of rectangular plaquettes,  topologically equivalent to a square lattice. There  we introduce the emergent vector field $\vec E$, as in the text. The circulation of the vector field  along  any closed loop  is zero. Middle and Bottom: The Shakti's low-energy manifold.  XMCD image of Shakti spin ice, for a spin ensemble with one excitation (red and blue dots) and  the corresponding emergent dimer cover representation. Now excitations  appear as multiple occupancy and/or diagonal dimers (Type~II$_2$s). $\vec E$ is no longer irrotational and its circulation defines the topological charge as $q=\frac{1}{4}\oint_{\gamma}\vec{E} \cdot d\vec{l}$. (Image copyright: Yuyang Lao.) }
\label{default}
\end{figure}

Beyond the standard dimer model, this picture can incorporate the low-energy excitations of Shakti ice as scramblings of the  cover. As Fig.~11 shows, above the ground state a frustrated plaquette (i.e. a node of the dimer lattice) can be dimerized three times instead of one (over-dimerization) by UVs, or also diagonally by a Type-II$_4$ or a Type-II$_2$  vertex. In the presence of such scramblings the emergent vector field $\vec E$ is not irrotational  anymore. Indeed its circulation around any topologically equivalent loop encircling a scrambling defines the quantized topological charge of the defect as $q=\frac{1}{4}\oint_{\gamma}\vec{E} \cdot d\vec{l}$ (Fig~11). Thus, the excitations of the Shakti ice manifold are topological charges, turning the discrete scalar field $h$ that defines its order into a multivalued phase.  

We have now the full picture: a topological phase, which cannot be explored from within, but only via  a discrete kinetics of excitations whose topological charge is conserved. This picture is emergent, and not at all evident from, or indeed reminiscent of, the original spin structure. It also has consequences for the kinetics, in terms of ergodicity breaking, non-equilibration, and glassiness, as it is typical of a topological state with topologically protected excitations, that cannot be reabsorbed into the manifold individually,  and which evolves via a discrete kinetics. All these issues are still to be investigated in full, as Shakti ice might provide the first artificial, controllable, modifiable and fully characterizable magnetic system which provides non-topographic vistas of ergodicity breaking and non-equilibration as consequences of a classical topological order.

\section{Dimensionality Reduction: Tetris Ice}

\begin{figure}[t!]
\includegraphics[width=.95\columnwidth]{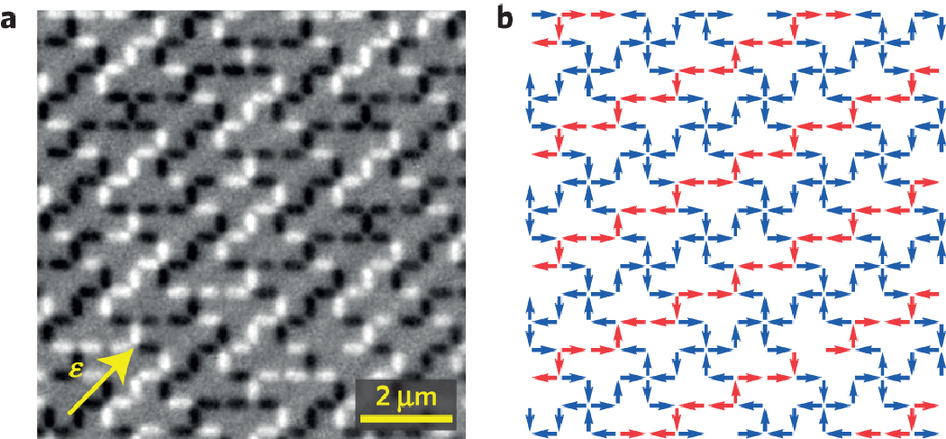}
\caption{Tetris Ice. (a) XMCD-PEEM image of a 600 nm Tetris lattice. The black/white contrast indicates whether the magnetization of an island has a component parallel or antiparallel to the polarization of the incident X-ray, which is indicated by the yellow arrow. (b) Map of the moment configurations  showing ordered backbones (blue) and disordered staircases (red). (Images from ref~\cite{gilbert2016emergent}.)}
\label{default}
\end{figure}

While Shakti spin ice provides a topologically protected low-energy manifold, no such protection is present in the  ground state of Tetris ice (Fig.~12), which can be explored by consecutive spin flips. As Fig.~12 shows, the lattice can be decomposed into T-shaped ``tetris'' pieces and it has a principal axis of symmetry.  

The geometry can be considered as layered one-dimensional systems. On the blue islands in Fig.~12 there cannot be any Type-II$_3$ unhappy vertex~\cite{Morrison2013}, and therefore the blue portion of the lattice, which we call backbone, must be ordered at the lowest energy. The unhappy vertices must reside on the red portions, which we call staircase, and which therefore remain disordered at low temperature. As temperature is lowered, we have thus a dimensional reduction of an alternating ordered-disordered one-dimensional system, which was indeed confirmed experimentally~\cite{gilbert2016emergent}.

This dimensional reduction is also apparent in the kinetics. Tetris was the first of the new geometries to be characterized in real-time, real-space and from the supplementary information of ref~\cite{gilbert2016emergent} it is possible to watch clips of its kinetics as the temperature is lowered or raised. Starting at high temperature, all the spins flip at about the same rate. As the temperature is lowered, ordered domains begin to form in correspondence with the 
backbones, where eventually the spins become static, while the spins on the staircases continue to fluctuate.

While Tetris spin ice' lowest energy state described above  has been confirmed experimentally, it follows from a nearest neighbor approximation. The profile of low energy  excitations, however, has not been yet studied in any systematic way, and promises  interesting new effects. For instance, as one-dimensional systems, the backbones can never order completely, and will always host excitations above the low-energy manifold. Of course Tetris is in fact a two-dimensional system, which decomposes into one-dimensional ones only in the lowest energy configuration. Slightly above such a manifold, one expects correlations among excitations that belong to different backbones. Such correlations must be controlled both by the magnetic interaction between these defects---as Tetris is, after all, a system of dipoles that can interact at long-range---but also through entropic interactions, since the backbones are separated by disordered staircases of non-zero density of entropy. None of the above issues has yet been studied theoretically, and they might indeed provide a useful setting to explore the  onset of phase decoupling  into lower-dimensional states, a broader problem relevant to liquid crystal phases~\cite{de1972tentative} or weakly coupled sliding phases~\cite{o1999sliding,sondhi2001sliding}.

\section{Polymers of Topologically Protected Excitations: Santa Fe Ice}

\begin{figure}[t!]
\includegraphics[width=.75\columnwidth]{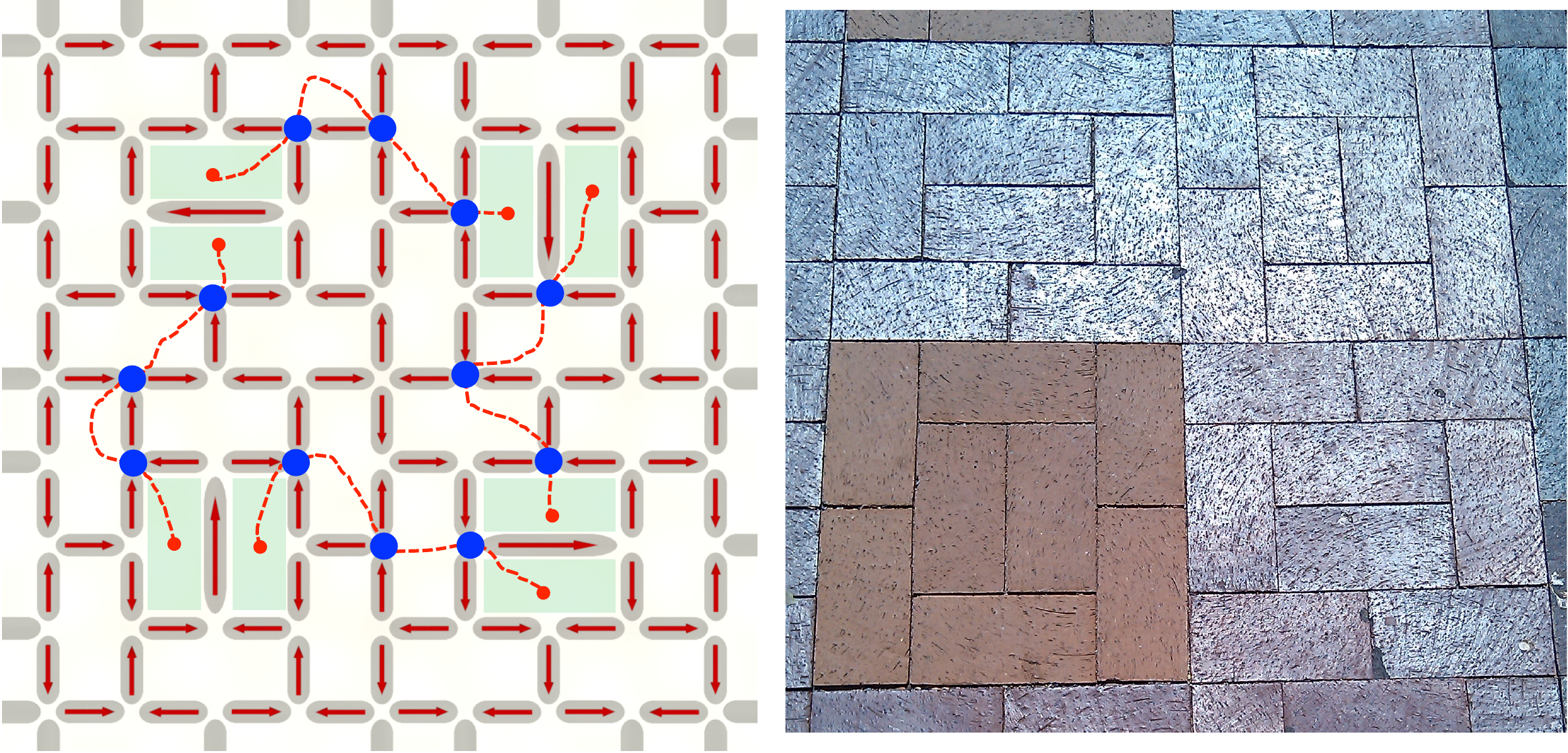}
\caption{The Santa Fe Ice can support both frustrated (shaded, green) and unfrustrated loops. There, ``polymers'' of unhappy vertices (blue dots) thread through unfrustrated loops to connect frustrated ones. On the right, the brick floor at the convention center in Santa Fe, New Mexico, USA.}
\label{default}
\end{figure}

We end this vista on how novel and unusual spin ice geometries influence topology with Santa Fe of Fig.~13, which was inspired by a terra cotta floor in the homonymous New Mexican capital---incidentally, the oldest in the United States. While Shakti and Tetris are 
maximally frustrated, which means that any minimal loop inside the geometry needs to be affected by an unhappy vertex, in Santa Fe only the dashed loops in the figure are frustrated and they are surrounded by unfrustrated ones. It is an inviolable topological constraint that at any energy frustrated loops can be affected by only an odd number of excitations, and unfrustrated ones  by only an even number (or none). 

An unhappy vertex on a frustrated loop of Santa Fe lattice affects a nearby unfrustrated one. However, an unfrustrated loop can only be affected by an even number of defects, and thus there will be a second unhappy vertex on it, affecting in turn a nearby unfrustrated loop, et cetera. It follows that magnetic ``polymers'', whose ``monomers'' are local protected excitations, must begin from and end into frustrated loops.

As each monomer costs energy, the lowest energy configuration of the magnetic ensemble will correspond to the shortest possible polymers, which are made of three monomers, each connecting nearby frustrated loops as in Fig.~13. The entropy of such a state can easily be  computed exactly. As each polymer dimerizes two frustrated loops, the degeneracy is given by the dimer cover model on the square lattice whose node is made of nearby frustrated loops, times the number of ways in which polymers can be chosen once their pinned ends are fixed. Thus the ice manifold decomposes into the direct product of two states: the dimer-cover manifold, which selects which loops are joined by the polymers, and  the  degeneracy of the polymers themselves. 

At low temperature the kinetics will reduce to the fluctuations of the magnetic polymers without changing the pinning location of their ends, and thus without changing the dimer cover picture. Thus, the low energy manifold can be explored from within, but only in part.  The kinetics within the manifold remains local and the polymer's fluctuations are  uncorrelated. As the temperature rises the polymers lengthen to include more than three monomers. At that point they can bump into each other, fuse in a cross, and then separate in different ways. This transition can lead to a different dimerization, as the new polymers, emerging from ``collisions'' of the old ones, are now pinned to different ending point. Thus the dimer-cover ensemble is explored via this mechanism of polymer colliding, fusing together  and then breaking again into different ones. This of course involves excitations over the ice manifold, further demonstrating the partial topological protection that pertains only to the dimer-cover sector of the low-energy manifold.

\section{Conclusions}

We have argued that by assembling together interacting, elementary building blocks, here Ising spins in the form of single domain, magnetic nano-islands, we can invert a tendency that has dominated condensed matter physics for half a century. Instead of finding serendipitously exotic states and behavior in nature, and then model them via higher level, emergent hamiltonians, one can devise materials---magnetic materials in this case--that do not exist in nature by developing first the model of their collective dynamics. Then, advances in nano--lithography and chemical synthesis can allow for their realization, while new thermal protocols afford characterization, often in real-time and real-space, for unprecedented direct validation of the theoretical expectations.    

We have shown that while the field began with simple geometries, reminiscent of natural materials, current advances make it possible to realize dedicated geometries of completely different properties and behaviors. These allow now to access rather sophisticated topological states, as their emergent description loses reminiscence of the original degrees of freedom, the underlying spin structure. This approach opens a new  path in the material-by-design effort, on  which unusual topological states can be deliberately designed. 

\bibliographystyle{unsrt}

\bibliography{library.bib}{}

\end{document}